\documentclass[12pt]{article}
\usepackage{cite}
\usepackage{amsmath, amssymb}
\usepackage{alltt}

\usepackage{subfig}
\usepackage{epsfig}

\usepackage{makeidx}         % allows index generation
\usepackage{graphicx}        % standard LaTeX graphics tool
\newtheorem{theorem}{Theorem}

\newcommand{\vect} \mathbf

\begin{document}

\title{Parallel random variates generator for GPUs based on normal numbers}
\author{Gleb Beliakov*, Michael Johnstone**, Doug Creighton**, Tim Wilkin*
\\
*School of Information Technology, \\Deakin University, \\221 Burwood Hwy, Burwood
3125, Australia \\ \texttt{gleb@deakin.edu.au, tim.wilkin@deakin.edu.au}
\\\\
**Centre for Intelligent Systems Research, \\Deakin University, \\Pigdons Rd, Geelong
3217, Australia
\\
}
\date{}
\maketitle

%\IEEEpeerreviewmaketitle

\begin{abstract}
Pseudorandom number generators are required for many computational tasks, such as stochastic modelling and simulation.
This paper investigates the serial CPU and parallel GPU implementation of a Linear Congruential Generator based on the binary representation of the normal number $\alpha_{2,3}$. We adapted two methods  of modular reduction which allowed us to perform most operations in 64-bit integer arithmetic, improving on the original implementation based on 106-bit double-double operations.
We found that our implementation is faster than existing methods in literature, and our generation rate is close to the limiting rate imposed by the efficiency of writing to a GPU's global memory.
\end{abstract}


\baselineskip=\normalbaselineskip

%\thispagestyle{empty}

%as many elements of $\vect x$ smaller than or equal to  $x_i$ as those bigger than or equal to $x_i$, provided $n$ is odd.

\section{Introduction}

Pseudorandom number generators are required for many computational tasks, such as simulating stochastic models, numerical integration and cryptography \cite{LEcuyer1994}. We focus on generation of random variates for numerical simulations, and hence do not consider generation of cryptographically strong random numbers, which are treated elsewhere \cite{Blum1982,HandbookAC1996}. For numerical simulations it is sufficient that the sequence of generated variates imitates a random sequence, even if the future and past elements can be predicted using the current element of the sequence. Typically a generator that produces uniform pseudorandom variates on $(0,1)$ (or random integers from $\{0,1,\ldots,m\}$) is the base generator used to generate random variates from more complicated distributions \cite{Atkinson1976,Hoermann2004,BeliakovCPC_2005}.

Linear Congruential Generators (LCG) and Mersenne Twisters (MT) are two of the most important families of random generators. MTs offer longer periods than LCGs and do not suffer from various correlations between the elements of the sequence as LCGs, especially when an LCG is poorly designed \cite{LEcuyer2002}. However MTs require a bigger state space and their implementation is more complex than that of LCGs.
Various statistical tests, such as SmallCrush and BigCrush from the TestU01 library \cite{LEcuyer2007} are used to evaluate the quality of the generated sequences.

General purpose Graphics Processing Units (GPU) have recently become a powerful alternative to traditional high performance computing. Certain calculations can be offloaded to the graphics card (referred to as GPU \emph{device}), on which thousands of threads are executed concurrently. GPUs offer hundreds of processor cores ($p=448$ in NVIDIA's Tesla C2050) at very modest prices, and are marketed as ``desktop supercomputers". GPUs are well suited for many numerical simulations, where algorithms offer a high degree of parallelism. Hence a significant interest to parallel pseudorandom variates generation, suitable for GPUs \cite{Howes2007,Saito2010,Passerat-Palmbach2011}. Important issues here are the speed of generation, the quality of the sequence generated, and the ability to produce the same sequence using different numbers of threads, number of hosts and GPU characteristics.

Recently Bailey and Borwein \cite{BaileySeptember2011} proposed a pseudorandom generator based on binary representation of normal numbers.
For an integer $b \geq2$ we say that a real number $K$ is $b$-normal  if, for all $m > 0$, every $m$-long string of digits in the base-$b$ expansion of $K$ appears, in the limit, with frequency $b^{-m}$.
That is, with exactly the frequency one would expect if the digits appeared completely ``at random". The authors presented a thorough theoretical overview of the topic, and designed a special class of constants $\alpha_{b,c}$, where $b,c>0$ and co-prime. They proved that such constants are $b$-normal, and therefore their base-$b$ expansion can serve to generate pseudorandom sequences.

A useful feature of the numbers $\alpha_{b,c}$ is that their  $n-$th base$-b$ digit can be calculated directly, without needing to compute any of the first $n - 1$ digits.
It implies that any element of the generated  sequence can be computed without computing the preceding elements, and therefore offers the desirable skip ahead feature of a random number generator.
It is valuable for parallel random number generators, as consecutive subsequences of  random variates can be generated concurrently on different processors or even different hosts, and the entire sequence is identical to the one generated in a single thread.

Bailey and Borwein \cite{BaileySeptember2011} present a pseudorandom number generator based on the binary representation of $\alpha_{2,3}$, and outline its implementation as a version of LCG, with the period $P=2 \cdot 3^{32} \approx 3.7 \cdot 10^{15}$. The seed of this generator is the index of the starting element, which is calculated directly through a simple formula. They suggest its use in parallel computations, by splitting the sequence of variates into $m$ subsequences ($m$ is the number of threads), each subsequence $i$ starting from the $1+\frac{(i-1)n}{m}$-th element, where $n$ is the desired number of random variates.

In this work we follow the path suggested by Bailey and Borwein and investigate a parallel implementation of their pseudorandom number generator suitable for GPUs.
Firstly we confirm the quality of the random sequence using the comprehensive suite BigCrush from the TestU01 library \cite{LEcuyer2007}.
Secondly we port the authors' implementation \cite{BaileySeptember2011} to GPU, and benchmark its speed. Initial investigations found that speed wise the proposed generator was comparable to but less efficient than the newest GPU-based MT generator \texttt{mtgp} \cite{Saito2010}. Therefore we investigate improvements to Bailey's implementation, which is based on either 106 bit double-double, or 128-bit integer arithmetic. We found several alternatives that use modular arithmetic, and improved the generation rate four-fold. Lastly we investigate the impact of GPU architecture on the generation rate, and determine an optimal configuration concerning the number of threads and thread blocks, and therefore work assigned to each thread. Our implementation delivers generation rate of 11 Gnumbers/sec on GPU, which is 25\% faster than state of the art \texttt{mtgp}, and approaches the maximal rate of 12.4 Gnumbers/sec imposed by the speed of writing to global memory.

The paper is structured as follows: Section \ref{sec1} reviews the LCG proposed by Bailey and Borwein. Section \ref{sec2} describes improvements that can be made to the original implementation using modular arithmetic. Numerical experiments are described and results presented in Section \ref{sec3}. Finally we draw conclusions in Section \ref{sec6}.

\section{Random number generator based on normal numbers}\label{sec1}

Bailey and Borwein \cite{BaileySeptember2011} investigated the following class of constants, which was introduced previously in \cite{Bailey2000},
$$
\alpha_{b,c}(r)=\sum_{k=1}^\infty \frac{1}{c^k b^{c^k+r_k}}
$$
where integers $b,c>1$ and co-prime and $r \in [0,1]$, and where $r_k$ denotes the $k$-th binary digit of $r$.

Bailey and Crandall \cite{Bailey2000} proved the following:

 \begin{theorem}
Every number $\alpha_{b,c}(r)$ is $b$-normal.
\end{theorem}

Bailey and Borwein focus on the number
$$
\alpha=\alpha_{2,3}(0) = \sum_{k=1}^\infty \frac{1}{3^k 2^{3^k}}
$$
This number is  2-normal, but is not 6-normal, as Bailey and Borwein show.  Further they show that all constants $\alpha_{b,c}$ are not $bc-$normal for co-prime integers $b,c > 1$.

The normality of $\alpha$ suggests that its binary expansion can be used to generate a sequence of pseudorandom numbers. Let $x_n=\{2^n \alpha\}$ be the binary expression of $\alpha$ starting from position $n+1$, where $\{\cdot\}$ denotes the fractional part of the argument. Bailey and Borwein show that when $n$ is not the power of three
\begin{equation} \label{eq1}
x_n= \frac{(2^{n-3^m} \lfloor3^m/2\rfloor) \mod 3^m}{3^m} + \varepsilon,
\end{equation}
and the tail of the series $\varepsilon < 10^{-30}$ if $n$ is not within 100 of any power of three.

Then the authors of  \cite{BaileySeptember2011} construct the following algorithm to calculate pseudorandom 64-bit real numbers in $(0,1)$, which would contain in their mantissas 53-bit segments of the binary expansion of $\alpha$
\\
\\
\textbf{Algorithm BB}

\begin{enumerate}
  \item Select starting index $a$ in the range $3^{33}+100$ to $2^{53}$, referred to as the seed of the generator.
  \item Calculate
        \begin{equation} \label{eq2}
            z_0=2^{a-3^{33}}\left\lfloor\frac{3^{33}}{2}\right\rfloor \mod 3^{33}
        \end{equation}
  \item Generate iterates
    \begin{equation} \label{eq3}
        z_k=2^{53}z_{k-1} \mod 3^{33}
    \end{equation}
    and return $z_k 3^{-33}$, which are 64 floating point random variates in $(0,1)$.
\end{enumerate}

Bailey and Borwein note that several operations need to be done with accuracy 106 mantissa bits, i.e., in 106-bit floating point arithmetic. They described and implemented their algorithm in double-double arithmetic \cite{BaileySeptember2011}. Note that double-double arithmetic allows exactly the representations of integers up to $2^{106}$.

The algorithm BB is a version of the LCG, and therefore it's period can be established from the theory of LCGs and results in $P=2 \cdot 3^{32} \approx 3.7 \cdot 10^{15}$. The algorithm can be checked by calculating $z_k$-th iterate recursively, or directly by using formula (\ref{eq2}). The binary digits of $\alpha$ within a range of indices spanned by successive powers of three are given by an LCG with a modulus that is a large power of three. Of course, by using a modulus, as large as $3^{40}$ one can get larger periods, but this will involve a higher number of bits in integer arithmetic, i.e. 128-bit integer arithmetic.

This approach has an advantage over LCGs that use a power of two value as the modulus, where arrays of pseudorandom data, of size matching a power of two, are accessed by row and column and therefore have the potential for a reduced period \cite{BaileySeptember2011}. A modulus as that proposed by Bailey and Borwein removes this issue.

The direct formula (\ref{eq2}) produces the skip ahead property: the ability to calculate the $k$-th iterate without iterating through the previous steps. This property is valuable for parallel random number generators, as every parallel thread can calculate its starting iterate directly from (\ref{eq2}).

\section{Parallelization and improvements to the algorithm}\label{sec2}

Using Bailey and Borwein's (BB) implementation \cite{BaileySeptember2011} as the starting point, we translated it from FORTRAN to C, and implemented as a parallel random number generator on GPU platform using CUDA \cite{NVIDIA2007}. Our initial tests confirmed that the pseudorandom sequence produced was identical to the one produced by using the original implementation in a single thread. Our generation rate was  0.6 Gnumbers per second on Tesla C2050 card.

%We obtained a substantial improvement of the generation rate from 0.028 (reported in \cite{}) to 0.9 Gnumbers per second on NVIDIA's Tesla C2050 card.

However, when compared to the state of the art MT generator developed by Saito and Matsumoto in 2010 \cite{Saito2010}, which has the rate of over 9 Gnumbers/sec for doubles in $(0,1)$ on the same card, the BB's implementation was not competitive. We note that the MT generator \texttt{mtgp} we benchmarked against has the periods from $2^{11213}-1$ to $2^{44497}-1$, and is free from the usual problems of LCG such as small periods and some statistical flaws.

An alternative implementation of the BB algorithm is to use 128 bit arithmetic. NVIDIA have released such a library for CUDA \cite{NVIDIA2012b}, available to registered developers. This library implements in software 128 bit arithmetic, as GPU hardware support for 128 bit arithmetic is not currently available. The generation rate was similar to that of the BB implementation, still uncompetitive with  \texttt{mtgp}.

Nevertheless we were still interested in the BB's LCG for two reasons. First, LCG's implementation is much simpler than that of MTs, and in particular the well optimized implementation from \cite{LEcuyer1988}. Simpler implementations are more portable and less tied to hardware characteristics than complex implementations such as \texttt{mtgp}. Second, BB's LCG offers an acceptable period, suitable for many simulation applications, which can also be easily extended using a larger modulus and therefore making the range between consecutive powers of three larger. Third, MT's state is larger than that of LCG, and Saito and Matsumoto's implementation involves cooperative generation of the sequence by several threads keeping the state in the shared memory. This could be inconvenient when using this generator in complex simulations on GPU, as pointed out by \cite{Gladkov2012}, which have their own demands on shared memory. In contrast,  BB's LCG offers fully independent generation of the sequence by different threads and only a 64-bit state.
Therefore we investigated possible improvements to the LCG.

Our focus was the generation step Eq. (\ref{eq3}). Eq. (\ref{eq2}) needs to be applied only once at the start of the process, and contributes to the total time only marginally, as long as the majority of the sequence is generated by Eq. (\ref{eq3}).

\subsection{Modular arithmetic}

A modular reduction is simply the computation of the remainder of an integer division. However, division is a much more expensive operation than multiplication. There are a number of special methods for performing modular reduction using multiplication, summation and subtraction, and  single precision division.

\subsubsection{L'Ecuyer method}

The first approach to modular reduction in LCG was adapted from P. L'Ecuyer \cite{LEcuyer1988}. This approach follows the earlier work \cite{Bratley1987}, and is based on modular multiplication $a z\mod m$, where $m=a q +r$, $r<a$ and $a^2<m$. One has
%For $x<m$ ???
$$
a z \mod m = (a(z \mod q) - \left\lfloor\frac{z}{q}\right\rfloor r) \mod m
$$

The algorithm is stated as follows.\\

\textbf{Algorithm L'Ecuyer}

Input: $z,a,m$, $q=\lfloor\frac{m}{a}\rfloor$, $r=m \mod a$

Output: $s=(a \cdot z) \mod m$

\begin{enumerate}
  \item $t= z \; div \;q$
  \item $s = a (s- t \cdot q) - t \cdot r$
  \item if($s<0$) then $s=s+m$
  \item return $s$
\end{enumerate}

The quantities $q=\lfloor\frac{m}{a}\rfloor$ and $r=m \mod a$ are pre-computed. Every intermediate step remains strictly between $-m$ and $m$ as shown in \cite{LEcuyer1988}, and hence only one \texttt{if} statement is needed to perform the modulus operation.

However we cannot use  L'Ecuyer's algorithm directly for calculating $2^{53}z_{k-1} \mod m$ because $a=2^{53}>3^{33}=m$, and the algorithm requires $a^2<m$. Instead we apply this reduction algorithm twice using $a=2^{25}$, and expanding
$$2^{53} z \mod m = 2^{25} \cdot 2(2^{25} \cdot 4 z \mod m) \mod m.$$ Hence we call L'Ecuyer's algorithm in the following sequence of calls

$z=LEcuyer(4z_{k-1},a,m,q,r)$

$z_k=LEcuyer(2z,a,m,q,r)$

Here all intermediate operations are less than 64-bits. Let us show that $s$ stays in $[-m,m]$. An essential condition guaranteeing
that the intermediate steps are between $-m$ and $m$, and hence correctness of $\mod m$ in step 3, is $z<m$ \cite{LEcuyer1988}. But in our case $z=4 z_{k-1} < 4 m$, so $z$ can be larger than $m$.

We modify the argument from \cite{LEcuyer1988} as follows.
$$
\left\lfloor \frac{z}{q}\right\rfloor r < 4 \left\lfloor \frac{a q +r}{q}\right\rfloor r  \leq 4 a r \leq 4 a^2 =  4 \cdot 2^{50} = 2^{52} < 3^{33}  = m.
$$
Therefore  $s$ at step 2 of L'Ecuyer's algorithm is  between $-m$ and $m$ for our specific values of $m$ and $a$, and hence the algorithm works correctly when $z\leq 4 m$.

We also note that computations are faster if instead of integer division in step 1 we multiply $z$ by the pre-computed reciprocal of $q$ (a 64-bit floating point number), which provides sufficient (53-bits) accuracy for our purposes, because its result is between $-m$ and $m$.
The actual C source code is presented in Figure \ref{figcodelcn}.

\begin{figure}[!h]
\renewcommand{\baselinestretch}{1}
 \begin{alltt}
\begin{minipage}{13cm}\small

\#define lcn(s, m, q, r, qinv)\textbackslash
    T1 = (s)*qinv;\textbackslash
    s = ((s-T1*q)<<25)-T1*r;\textbackslash
    if (s < 0)	s+=m;
		
\#define lcn4(s, m, q, r, qinv)\textbackslash
    T1 = (s)*qinv;\textbackslash
    s =  (((s)-T1*q)<<25 )- T1*r;\textbackslash
    s+=m;

...
for(int i = 0; i < WorkPerThread; i++)
\{
  s<<=2;
  lcn4(s, m, q, r, qinv);
  s<<=1;
  lcn(s, m, q, r, qinv);
  output[startindex + i] = r3i  * s;
\}
\end{minipage}
\end{alltt}
\caption{A fragment of our implementation of the modified L'Ecuyer's method in C (as a macro).
}\label{figcodelcn}
\renewcommand{\baselinestretch}{2}
\end{figure}

\subsubsection{Barrett's reduction}

Next, we looked at Barrett's reduction \cite{Barrett1987}.
Barrett introduced the idea of estimating the quotient $\lfloor \frac{x}{m}\rfloor$ with operations that are either less expensive in time than a multi-precision division by $m$ or can be done as a pre-calculation for a given $m$. Barrett uses the approximation
$$
\left\lfloor \frac{x}{m}\right\rfloor -2 \leq \left\lfloor \frac{\left\lfloor\frac{x}{2^{k-1}}\right\rfloor \cdot \left\lfloor \frac{2^{k-1}{2^{k+1}}}{m} \right\rfloor}{2^{k+1}}\right\rfloor=\left\lfloor \frac{\left\lfloor\frac{x}{2^{k-1}}\right\rfloor \cdot \left\lfloor \frac{2^{2k}}{m}\right\rfloor}{2^{k+1}}\right\rfloor \leq \left\lfloor \frac{x}{m}\right\rfloor
$$
and the equation $x \mod m =x-m \left\lfloor \frac{x}{m}\right\rfloor$.
The term $\lfloor \frac{2^{2k}}{m}\rfloor$ depends only on $m$ and can be pre-computed. The other divisions are efficient binary shift operations. The value of $k$ here is the number of binary digits of $m$. The algorithm reads as follows\\

\textbf{Algorithm Barrett}

Input: $x=(x_{2k-1}\ldots x_1x_0)_2$,$m=(m_{k-1}\ldots,m_1m_0)_2$, $\mu=\left\lfloor \frac{2^{2k}}{m}\right\rfloor$, $x,m>0$, $m_{k-1}\neq 0$

Output: $r= x \mod m$
\begin{enumerate}
  \item $q_1=\left\lfloor \frac{x}{2^{k-1}}\right\rfloor$, $q_2=\mu \cdot q_1$, $q_3=\left\lfloor \frac{q_2}{2^{k+1}}\right\rfloor$
  \item $r_1=x \mod 2^{k+1}$, $r_2=q_3 \cdot m \mod 2^{k+1}$, $r=r_1-r_2$
  \item if $r<0$  then $r=r+2^{k+1}$
  \item while $r \geq m$ do $r=r-m$
  \item return $r$.
\end{enumerate}

In our case  $k=53$, and  $\mu$ is pre-computed in extended arithmetic but is a 64-bit integer itself. Integer $q_2$ needs extended precision, but $q_1$ and $q_3$ need only 64 bits using binary shift operations with high and low parts of $q_2$. Integer $q_3\cdot m$ requires extended arithmetic. Finally $r_1, r_2$  and $r$ are 64 bit integers. These observations allow one to reduce CPU time. At most two subtractions at step 4 are needed. The actual C code is presented in Figure \ref{figcodebarrett}.

\begin{figure}[!h]
\renewcommand{\baselinestretch}{1}
 \begin{alltt}
\begin{minipage}{13cm}\small

\#define barrett_step_simple(rlo)\textbackslash
  xlo = t53 * rlo;\textbackslash
  xhi = __umul64hi (t53, rlo);\textbackslash
  qlo = (xhi << 12) | (xlo >> 52);\textbackslash
  qhi = __umul64hi(qlo, mulo);\textbackslash
  qlo = qlo * mulo;\textbackslash
  qlo = (qhi << 10) | (qlo >> 54);\textbackslash
  qhi = 0;\textbackslash
  r1lo = xlo & 0x3FFFFFFFFFFFFFULL;\textbackslash
  r2lo = qlo * m;\textbackslash
  r2lo = r2lo & 0x3FFFFFFFFFFFFFULL;\textbackslash
  rlo = r1lo - r2lo;\textbackslash
  if (r1lo<r2lo) rlo += 0x40000000000000ULL;\textbackslash
  while (rlo >= m) rlo -= m;
...
for(int i = 0; i < WorkPerThread; i++)
\{
  barrett_step_simple(rlo);	
  output[startindex + i] = r3i * rlo;
\}
\end{minipage}
\end{alltt}
\caption{A fragment of our implementation of the Barrett's method in C (as a macro). Note that we use CUDA's operation \texttt{\_\_umul64hi} which returns the most significant 64 bits of the product of two 64 bit integers.
}\label{figcodebarrett}
\renewcommand{\baselinestretch}{2}
\end{figure}

\subsubsection{Modified Barrett's reduction}

We note that with our choice of $\mu=\left\lfloor \frac{2^{2k}}{m}\right\rfloor$, at step 1 of Barrett's algorithm we have $q_1=\left\lfloor \frac{x}{2^{53-1}}\right\rfloor = \left\lfloor \frac{2^{53} z}{2^{52}} \right\rfloor = 2 z$, and consequently $q_2=2 \mu z$. Similarly at Step 2 we have $r_1=2^{53}z \mod 2^{52} = z \mod 2$.
We questioned whether these two operations can be eliminated altogether.

We rewrite Barrett's formula as follows
$$
 \left\lfloor \frac{\left\lfloor\frac{x}{2^{k}}\right\rfloor \cdot \left\lfloor \frac{2^{k}{2^{k}}}{m} \right\rfloor}{2^{k}}\right\rfloor=\left\lfloor \frac{\left\lfloor\frac{x}{2^{k}}\right\rfloor \cdot \left\lfloor \frac{2^{2k}}{m}\right\rfloor}{2^{k}}\right\rfloor \leq \left\lfloor \frac{x}{m}\right\rfloor.
$$
Then we can keep the same $\mu=\left\lfloor \frac{2^{2k}}{m}\right\rfloor$, but the two mentioned operations simplify as follows:

1) at Step 1 we calculate $q_1=\left\lfloor \frac{x}{2^{53}}\right\rfloor = \left\lfloor \frac{2^{53} z}{2^{53}} \right\rfloor = z$;

2) at Step 2 we have $r_1=2^{53}z \mod 2^{53} = 0$.

Therefore these two instructions become redundant, and hence we save on CPU time. Furthermore, since $r=0-r_2<0$, the \texttt{if} statement in Step 3 is redundant. Finally, because in our case $3^{33}=m<2^{k}=2^{53}<2 m$, and because $r<2^{53}$, we have $r<2m$ and therefore \texttt{while} in Step 4 can be replaced with a cheaper \texttt{if}.
 Our modified Barrett's algorithm becomes\\

\textbf{Algorithm Modified Barrett}

Input: $z=(z_{k-1}\ldots z_1z_0)_2$,$m=(m_{k-1}\ldots,m_1m_0)_2$, $\mu=\left\lfloor \frac{2^{2k}}{m}\right\rfloor$, $x,m>0$, $m_{k-1}\neq 0$

Output: $r= 2^{k} z \mod m$
\begin{enumerate}
  \item  $q_2=\mu \cdot z$, $q_3=\left\lfloor \frac{q_2}{2^{k}}\right\rfloor$
  \item $r_2=q_3 \cdot m \mod 2^{k}$
  \item $r=2^{k} - r_2$
  \item if $r \geq m$ then $r=r-m$
  \item return $r$.
\end{enumerate}

Our C implementation of this algorithm is shown in Figure \ref{figcodemodbarrett}.

\begin{figure}[!h]
\renewcommand{\baselinestretch}{1}
 \begin{alltt}
\begin{minipage}{13cm}\small
#define stepBarrett(z)\textbackslash
  qhi =__umul64hi(z, mu);\textbackslash   /* high part of $q_2$ */
  z *=  mu;\textbackslash                 /* low part of $q_2$ */
  z = ( ((qhi << 11) | (z >> 53) )* m )\&0x1FFFFFFFFFFFFFULL;	\textbackslash
  z= 0x20000000000000ULL - z; \textbackslash  /* $2^53 - r_2$ */
  if (z >= m) z -= m;
...
for(int i = 0; i < WorkPerThread; i++)
\{	
  stepBarrett(z);
  output[startindex + i] = r3i * z;
\}
\end{minipage}
\end{alltt}
\caption{A fragment of our implementation of the modified Barrett's method in C (as a macro). Note that
 x \&0x1FFFFFFFFFFFFFULL implements $x \mod 2^{53}$ operation and the shifts implement $q_2/2^{53}$. Operation \texttt{\_\_umul64hi} which returns the most significant 64 bits of the product of two 64 bit integers
}\label{figcodemodbarrett}
\renewcommand{\baselinestretch}{2}
\end{figure}

The proposed three methods of reduction have a generation rate of 0.6 Gnumbers/sec, and still did not match the rate provided by \texttt{mtgp}. Therefore we also attempted to accelerate calculations using programming techniques. Our first suspicion was the speed of writing to GPU's global memory.

\subsection{Writing to global memory}

It is known that both global and shared memory access pattern affects the performance of GPU algorithms \cite{NVIDIA2012}. Strided access, when the values are read from or written to GPU memory by each thread not consecutively but with certain steps, helps avoid bank conflicts and execute read/write operations more efficiently.
While on many occasions the generated random numbers need not be stored in memory but rather used in a simulation routine on GPU, it is helpful to measure  the generation rate when the random numbers are written into an array in the global memory. It allows one to benchmark the algorithm against the competitors in a fair way, and it also avoids the possibility that some instructions can be skipped by the optimizer if their results are not used. Furthermore, knowing the speed at which an algorithm writes arbitrary fixed values into global memory under the same memory access pattern, one can calculate the actual net generation speed.

For these reasons we also experimented with different steps at which threads write consecutively generated random numbers into global memory, as a function of the number of threads and thread blocks. As expected, letting each thread to write into consecutive positions in the output array  was less efficient, because on GPUs read/writes to global memory are performed in blocks (of 128 bytes, as specified in CUDA Programmer Manual \cite{NVIDIA2012}). Therefore when $p$  concurrent threads write to consecutive positions, they require just one block write for all $p$ generated values (if they all fit one block), as opposed to $p$ writes when they write to non-coalescent positions.

While the elements of the generated sequence will not be located in the global memory consecutively, this is not an issue, because the same access pattern can be applied when they need to be retrieved by a simulation algorithm (which is a more efficient pattern), therefore the sequence can be easily restored to its original order.

Our algorithm involves a parameter ``step" (see Figure \ref{figcodeunroll}), calculated based on the number of threads and thread blocks.  Our optimal configuration on Tesla C2050 involved 512 threads per block, and 4096 elements per thread, and therefore $\lceil \frac{n}{512 \cdot 4096}\rceil$ thread blocks. Thus $step$ was $512 \cdot blocks$.

Coalesced memory access  improved the generation rate by a factor of more than ten.

\subsection{Loop unrolling and inlining}

Loop unrolling is a standard method for improving numerical performance. Loop unrolling allows for more computation to be performed without the overhead of managing loop counters and checking ending conditions, but comes at the expense of additional register usage. We unrolled the generation loop in each thread as in Figure \ref{figcodeunroll}. Unrolling the loop for Barrett's method improved generation rate by 4.9\%, and the improvement to other methods was between 1 and 4\%.

\begin{figure}[!h]
\renewcommand{\baselinestretch}{1}
 \begin{alltt}
\begin{minipage}{13cm}\small
double r3i=1.0/5559060566555523.0;  // pow(3.0, -33.0);
unsigned long long mu = 0x33D9481681D79DULL; // pow(3.0, 33.0) div pow(2,53);
/* Calculate seed z using Eq. (\ref{eq2}).*/
for(unsigned int i = 0; i < WorkPerThread; i+=8)
\{
    stepBarrett(z);
    output[startindex + i*step] = r3i * z;
    stepBarrett(z);
    output[startindex + i*step+step] = r3i * z;
    stepBarrett(z);
    output[startindex + i*step+step*2] = r3i * z;
    stepBarrett(z);
    output[startindex + i*step+step*3] = r3i * z;
    stepBarrett(z);
    output[startindex + i*step+step*4] = r3i * z;
    stepBarrett(z);
    output[startindex + i*step+step*5] = r3i * z;
    stepBarrett(z);
    output[startindex + i*step+step*6] = r3i * z;
    stepBarrett(z);
    output[startindex + i*step+step*7] = r3i * z;
\}
\end{minipage}
\end{alltt}
\caption{Partial loop unrolling for modified Barrett's method.
}\label{figcodeunroll}
\renewcommand{\baselinestretch}{2}
\end{figure}

Another strategy is function inlining. It reduces computation time as no parameters need to go through stack. One can inline functions by using macros or inline directive. On GPU the device functions are automatically inlined, and we noticed only marginal benefit in using macro as opposed to inline function in our implementation of Barrett' algorithm.

\section{Numerical experiments}\label{sec3}
The two aspects of the generator to consider are the statistical quality and the generation rate, presented in the following sections respectively.

\subsection{Statistical testing}

The testing suite BigCrush \cite{LEcuyer2007} was applied to the generator algorithm BB. Of the 106 statistical tests, all tests passed except for the BirthdaySpacings tests \#13-21 and the ClosePairs tests \#22-24.

The Birthday Spacings tests assume the generation of a value of length 64bit. The BB algorithm generates a random value of length 53bits, which corresponds to the mantissa of a double value type. The restrictions placed on the parameters of the Birthday Spacings tests of BigCrush do not hold when considering only 53bits, therefore the failure of these tests is expected.

The Close Paris tests, as studied in \cite{LEcuyer2000}, would expect an LCG to fail with period less than $2^{60}$. Here we have a period of $P=2 \cdot 3^{32}$, and a failed test. The rest of the tests (94 in total) were passed, and we conclude that the generated sequence is of a good statistical quality.

\subsection{Numerical testing}
Our numerical experiments were performed with an Intel® Core™ i7-860 processor workstation with 4 GB RAM clocked at 2.8 GHz running Linux (Fedora 12), and with a Tesla C2050 GPU with 448 cores, 3 GB of global memory, and clocked at 1.15 GHz.
We selected our version of Bailey and Borwein's algorithm \cite{BaileySeptember2011} referred to as BCN, its 128-bit arithmetic implementation BCN128bit, the LCG adapted from L'Ecuyer \cite{LEcuyer1988} LCN, Barrett's reduction \cite{Barrett1987}, and a modified version of Barrett's, refer to sections \ref{sec1} and \ref{sec2}, for analysis. These methods were compared with a serial implementation of each, run on a CPU, and also compared with parallel generators,  \texttt{mtgp} \cite{Saito2010} and the CUDA-SDK Mersenne twister \cite{NVIDIA2012a}.

Each kernel was timed for 100 iterations for varying threads per block counts, ranging from 128 threads per block to 512 threads per block, with a step of 32 threads. The timing information for the best thread count was recorded. The need for different thread counts is attributed to the varying register demands of each kernel. The unrolled kernels in particular require more registers, affecting the occupancy and therefore can be tuned by varying the thread count.

The generation rate for the various methods, averaged over 100 runs, is given in Table \ref{tablerates}. The table includes the rate at which numbers are generated with and without the time required to setup the generator, for example the time required to generate a starting seed for each thread in the BCN methods. The table shows the increase of the generation rate as techniques to speed up CUDA were applied, such as coalesced memory access and unrolling.

\begin{table}[htbp]
  \centering
  \caption{The generation rate for the different methods proposed.}
    \begin{tabular}{l|r|r}
    Method &  Execute Rate GNum/sec & Including Setup GNum/sec \\
    \hline \hline
    BCN   & 0.6706 & 0.6703 \\
    BCNCoalesced & 9.1744 & 9.0347 \\
    BCNUnrolled & 9.4756 & 9.3438 \\
    BCN128bit & 0.6584 & 0.6582\\
    BCN128bitCoalesced & 3.8914 & 3.8690 \\
    BCN128bitUnrolled & 3.9311 & 3.9140\\
    LCN   & 0.6592 & 0.6590\\
    LCNCoalesced & 7.5974 &7.5015 \\
    LCNUnrolled &8.1274 &8.0422\\
    Barrett & 0.6660 & 0.6657 \\
    BarrettCoalesced & 8.0413 & 7.9337 \\
    BarrettUnrolled & 8.4334 & 8.3415 \\
    BarrettModified & 11.4434 & 11.2268 \\
     \hline
    Constant & 0.6900 & 0.6897\\
    ConstantCoalesced & 12.4127 & 12.2891 \\
    ConstantUnrolled & 12.6217 & 12.3587\\
     \hline
    mtgp  & 9.8455 & 9.5865 \\
    sdk MT & 4.2408 & 4.0872 \\
    \hline
    BCNSerial & -& 0.024 \\
    LCNSerial & -& 0.051 \\
    BarrettModifiedSerial & -& 0.102 \\
    ConstantSerial & -& 0.526 \\
    \end{tabular}%
  \label{tablerates}%
\end{table}%

The Constant methods in the table refer to a simple CUDA implementation where a single constant value is written back to global memory. This method provides a maximum generation rate that we are able to obtain. Since the Constant methods use the same memory access as the random number generators, they describe the maximum rate achievable due to memory access bottlenecks.

In Table \ref{tablerates} it is evident that coalesced memory access increases the generation rate of the kernels, this is especially true in the case of memory bound algorithms such as the ones under analysis here. The improvements unrolling can make can also be seen, however this is a slight improvement when compared to the coalesced improvement.

From Table \ref{tablerates} it can be seen that the CPU-based serial methods are inferior to those of the generators running in parallel in CUDA. While global memory access is expensive  in CUDA, on CPU it does not represent a bottleneck. Indeed, ConstantSerial method  entry shows a very high memory write rate, which indicates that contribution due to memory access on CPU is negligible, and the serial algorithm is CPU bound. In contrast, the GPU algorithm is clearly memory bound.

The generation rate of the BarrettModified method can be seen to approach that of the Constant method. If we remove the memory access bottleneck from the timing results, as seen in Table \ref{tabletime}, an appreciation of the net generation rate without expensive memory access can be made. Here the parallel optimized Barrett method exceeds the serial implementation by approximately 19\%.

\begin{table}[htbp]
  \centering
  \caption{The generation rate for various methods with memory access bottlenecks removed.}
    \begin{tabular}{l|r| r| r}
    Method & RunTime (ms) & Generation Time (ms) & Generation Rate (GNum/sec) \\
    \hline \hline
    %BCN   & 400.427865 & 11.23859 & 23.88515 \\
   % BCNCoalesced & 29.716942 & 7.873218 & 34.09476 \\
    BCNUnrolled & 28.7293 & 7.0049& 38.3206 \\
    %BCN128bit & 407.83974 & 18.65047 & 14.39296 \\
    %BCN128bitCoalesced & 69.381694 & 47.53797 & 5.646759 \\
    BCN128bitUnrolled & 68.5847 & 46.8604 & 5.7284 \\
   % LCN   & 407.32974 & 18.14047 & 14.7976 \\
    %LCNCoalesced & 41.452101 & 19.60838 & 13.68984 \\
    LCNUnrolled & 33.3808 & 11.6565 & 23.0287 \\
   % Barrett & 403.203568 & 14.0143 & 19.1544 \\
   % BarrettCoalesced & 33.840648 & 11.99692 & 22.37536 \\
    BarrettUnrolled & 32.1831 & 10.4588 & 25.6658 \\
    BarrettOpt & 23.9146 & 2.1902 & 122.5574 \\
 \hline
   % Constant & 389.189271 &       & 0.68973 \\
    %ConstantCoalesced & 21.843724 &       & 12.28891 \\
    ConstantUnrolled & 21.7243 & -     & 12.3564 \\
    \end{tabular}%
  \label{tabletime}%
\end{table}%

\subsection{Serial methods under different compilers}

Our GPU parallel implementation of the Barrett's method involves the function \texttt{\_\_umul64hi} as part of one 128-bit multiplication, see Figure \ref{figcodemodbarrett}. This is CUDA's compiler \texttt{nvcc} intrinsic function, which is supported only on the GPU device. We experimented with several alternative implementations for CPU, which depend on both the compiler and hardware.

There is a generic implementation of this function available from \texttt{http://opencl-usu-2009.}\\\texttt{googlecode.com/svn/trunk/inc/dynlink/device\_functions\_dynlink.h},
presented in Figure \ref{fig_mul64}.

Microsoft Visual Studio offers the intrinsic function \texttt{\_umul128} which performs 128-bit multiplication of two 64-bit unsigned integers and returns the high and low bits of the result. Gcc compiler on Linux (64-bit version only) offers the function \texttt{mul128} which returns the 128-bits result (data type \texttt{\_\_uint128\_t}). This function in particular is translated into just three assembler instructions on Intel 64-bit processors, and is therefore extremely efficient. The usage of these functions is also shown in Figure \ref{fig_mul64}.

Therefore we expect that the efficiency of the generation on CPU will depend on the hardware and compiler being used, and also the compilation parameters. In Table \ref{tabletime} we present generation rates on different machines under different compilers. What is noticeable is that the original BCN implementation performs quite well in some cases, although on 64 bit Intel architecture our modified Barrett method still delivers the highest generation rate. It is interesting that modified L'Ecuyer's method produced the highest generation rate in the debug mode (i.e., no compiler optimizations), while it lags behind in the release mode.

For comparison, we also timed the standard C function \texttt{rand} (often criticized for its low statistical quality) by calculating uniform pseudorandom numbers from (0,1) using the following instructions:

\texttt{const double RAND\_MAX\_INV=1.0/(RAND\_MAX+1.0);}

\texttt{rand()*RAND\_MAX\_INV;}

We found that the generation rate of the \texttt{rand} function is less than half of
the rate of the Barrett method, and is comparable to the L'Ecuyer's method.

The speedup offered by the GPU is noticeable from Tables \ref{tabletime}-\ref{tabletimeserial},  and when excluding access to GPU global memory, is more than 1000-fold.

\begin{table}[htbp]
  \centering
  \caption{The  generation rate for various methods on CPU using different compilers (Gnum/sec). }
    \begin{tabular}{l|r|r|r|r|r|r}
    Method &  gcc 4.0  &  gcc 4.0  & VC 64 bit & VC 64 bit & VC 32 bit & VC 32 bit\\
     &   -m64 -O3 &  -m64 -g & release  & debug & release  & debug\\
    \hline \hline
    BCN   & 0.024 & 0.014 & 0.025 & 0.012 & 0.026 & 0.0084  \\
    LCN   & 0.051 & 0.032 & 0.060 & 0.036 & 0.030 & 0.013  \\
    Barrett & 0.102 & 0.024 & 0.067 & 0.020 & 0.026 & 0.076  \\
    BarrettModified & 0.102 & 0.045 & 0.073 & 0.029 &  0.025 & 0.079  \\
    Rand & 0.051 & 0.051 & 0.063 & 0.063 &  0.030 & 0.030  \\
     \hline
    \end{tabular}%
  \label{tabletimeserial}%
\end{table}%

\begin{figure}[!h]
\renewcommand{\baselinestretch}{1}
 \begin{alltt}
\begin{minipage}{13cm}\small
/* */
typedef ULLong unsigned long long int;
inline unsigned int __umulhi(unsigned int a, unsigned int b)
\{
  ULLong c = (ULLong)a * (ULLong)b;
  return (unsigned int)(c >> 32);
\}

inline ULLong __umul64hi(ULLong a, ULLong b)
\{
  unsigned int           a_lo = (unsigned int)a;
  ULLong a_hi = a >> 32;
  unsigned int           b_lo = (unsigned int)b;
  ULLong b_hi = b >> 32;
  ULLong m1 = a_lo * b_hi;
  ULLong m2 = a_hi * b_lo;
  unsigned int           carry;
  carry = (0ULL + __umulhi(a_lo, b_lo) + (unsigned int)m1 + (unsigned int)m2) >> 32;
  return a_hi * b_hi + (m1 >> 32) + (m2 >> 32) + carry;
\}

/* This version is for Microsoft VC 64 bit compiler only*/
#include <xmmintrin.h>
#pragma intrinsic(_umul128)

#define stepBarrett(z)\textbackslash
   qlo=_umul128(z,mu,&qhi);\textbackslash
   r2lo = (((qhi << 11) | (qlo >> 53)) * m)& 0x1FFFFFFFFFFFFFULL;\textbackslash
   z = 0x20000000000000ULL - r2lo;\textbackslash
   if (z >= m) z -= m;

/* This version is for gcc 64 bit compiler only */
__uint128_t qq;   uint64_t z,mu,m;
typedef __uint128_t u128b __attribute__((mode(TI)));
__uint128_t mul128(__uint64_t a, __uint64_t b)
\{ return (__uint128_t) a * b; \}
#define stepBarrett(z)\textbackslash
   qq=mul128(z,mu);\textbackslash
   qq>>=53;\textbackslash
   r2lo = (qq * m)& 0x1FFFFFFFFFFFFFULL;\textbackslash
   z = 0x20000000000000ULL - r2lo;\textbackslash
   if (z >= m) z -= m;

\end{minipage}
\end{alltt}
\caption{Various implementations of the 128-bit multiplication.
}\label{fig_mul64}
\renewcommand{\baselinestretch}{2}
\end{figure}

%and note whether Bigcrush was passed or not (copy from MT paper)

\section{Conclusion}\label{sec6}

We presented a pseudorandom number generation algorithm based on  Bailey and Borwein's work on normal numbers and their original implementation as a version of LCG. We confirmed that the generated sequence is statistically suitable for simulation purposes by running a comprehensive BigCrush suite of tests. We improved Bailey and Borwein's implementation by implementing a special modular reduction algorithm, which gave speedup of the factor of 4 on a CPU. Our generator is twice as fast as the standard C function \texttt{rand()}. 

Further, we implemented a parallel version of the proposed algorithm suitable for GPUs and benchmarked it against the state-of-the-art Mersenne Twister parallel generator \texttt{mtgp}. We found that our implementation is faster than \texttt{mtgp}, and our generation rate is close to
the limiting rate imposed by the efficiency of writing to GPU's global memory. While MT's period is much longer than that of LCG, our LCG's advantage over \texttt{mtgp} is the simplicity and portability of implementation, independent generation of subsequences and less demand on GPU's scarce shared memory. We achieved speedup of 1000 times over the performance of random variate generator on CPU.
Our implementation is available from \texttt{http://www.deakin.edu.au/$\sim$gleb/bcn\_random.html}.

\bibliographystyle{plain}
\bibliography{median5}
\end{document}